\documentclass[runningheads]{llncs}

\usepackage{amsmath,amssymb,amsfonts}
\usepackage{array}
\usepackage{balance}
\usepackage{booktabs}
\usepackage{bm}
\usepackage{color,xcolor}
\usepackage{enumitem}
\usepackage{epstopdf}
\usepackage{graphicx}
\usepackage{multirow}
\usepackage{subfigure}
\usepackage{threeparttable}
\usepackage{textcomp}
\usepackage[colorlinks,linkcolor=blue,citecolor=blue,urlcolor=blue]{hyperref}

\makeatletter
\def\UrlAlphabet{%
      \do\a\do\b\do\c\do\d\do\e\do\f\do\g\do\h\do\i\do\j%
      \do\k\do\l\do\m\do\n\do\o\do\p\do\q\do\r\do\s\do\t%
      \do\u\do\v\do\w\do\x\do\y\do\z\do\A\do\B\do\C\do\D%
      \do\E\do\F\do\G\do\H\do\I\do\J\do\K\do\L\do\M\do\N%
      \do\O\do\P\do\Q\do\R\do\S\do\T\do\U\do\V\do\W\do\X%
      \do\Y\do\Z}
\def\UrlDigits{\do\1\do\2\do\3\do\4\do\5\do\6\do\7\do\8\do\9\do\0}
\g@addto@macro{\UrlBreaks}{\UrlOrds}
\g@addto@macro{\UrlBreaks}{\UrlAlphabet}
\g@addto@macro{\UrlBreaks}{\UrlDigits}
\makeatother

\renewcommand{\paragraph}[1]{\vspace{0.5em}\noindent\textbf{#1.}}
\renewcommand{\subparagraph}[1]{\vspace{0.5em}\noindent\textit{\underline{#1.}}}

\newcommand{\norm}[1]{\Vert #1 \Vert}
\newcommand{\num}[1]{\vert #1 \vert}

\makeatletter
\newif\if@restonecol
\makeatother

\usepackage[linesnumbered,ruled,vlined]{algorithm2e}
\usepackage{algpseudocode}

\SetKwComment{Comment}{$\triangleright$\ }{}

\begin{document}

%%%%%%%%%%%%%%%%%%%%%%%%%%%%%%%%%%%%%%%%%%%%%%%%%%%%%%%%%%%%%%%%%%%%%%%
%% Title
%%%%%%%%%%%%%%%%%%%%%%%%%%%%%%%%%%%%%%%%%%%%%%%%%%%%%%%%%%%%%%%%%%%%%%%
\title{DIOT: Detecting Implicit Obstacles from Trajectories}

\author{
Yifan~Lei\inst{1,2} \and
Qiang~Huang\inst{1}\thanks{Corresponding Author} \and %
Mohan~Kankanhalli\inst{1} \and
Anthony~Tung\inst{1}
}%
\authorrunning{Y. Lei et al.}

\institute{School of Computing, National University of Singapore, Singapore \and Tencent Inc., Shenzhen, China \\
\email{\{leiyifan,huangq,mohan,atung\}@comp.nus.edu.sg}}

\maketitle           % typeset the header of the contribution

%%%%%%%%%%%%%%%%%%%%%%%%%%%%%%%%%%%%%%%%%%%%%%%%%%%%%%%%%%%%%%%%%%%%%%%
%% Abstract
%%%%%%%%%%%%%%%%%%%%%%%%%%%%%%%%%%%%%%%%%%%%%%%%%%%%%%%%%%%%%%%%%%%%%%%
\begin{abstract}
In this paper, we study a new data mining problem of obstacle detection from trajectory data. 
Intuitively, given two kinds of trajectories, i.e., reference and query trajectories, the obstacle is a region such that most query trajectories need to bypass this region, whereas the reference trajectories can go through as usual. 
We introduce a density-based definition for the obstacle based on a new normalized Dynamic Time Warping (nDTW) distance and the density functions tailored for the sub-trajectories to estimate the density variations. 
With this definition, we introduce a novel framework \textsf{DIOT} that utilizes the depth-first search method to detect implicit obstacles. 
We conduct extensive experiments over two real-life data sets. The experimental results show that \textsf{DIOT} can capture the nature of obstacles yet detect the implicit obstacles efficiently and effectively. Code is available at \url{https://github.com/1flei/obstacle}.

\end{abstract}

%%%%%%%%%%%%%%%%%%%%%%%%%%%%%%%%%%%%%%%%%%%%%%%%%%%%%%%%%%%%%%%%%%%%%%%
%% Introduction
%%%%%%%%%%%%%%%%%%%%%%%%%%%%%%%%%%%%%%%%%%%%%%%%%%%%%%%%%%%%%%%%%%%%%%%
\section{Introduction}
\label{sect:intro}
With the prevalence of location devices, massive trajectory data have been generated and used for data analytics. The trajectory is a sequence of geo-locations of moving objects such as cars, vessels, and anonymous persons. In this paper, we study a new data mining problem of obstacle detection based on trajectory data. Given two kinds of trajectories, i.e., reference and query trajectories, the obstacle is a region such that most query trajectories need to bypass this region; in contrast, the reference trajectories go through as usual.

\begin{figure}[t]
\centering
\subfigure[Vessel Trajectories in May 2017]{
	\label{fig:example:query}
	\includegraphics[width=0.47\textwidth]{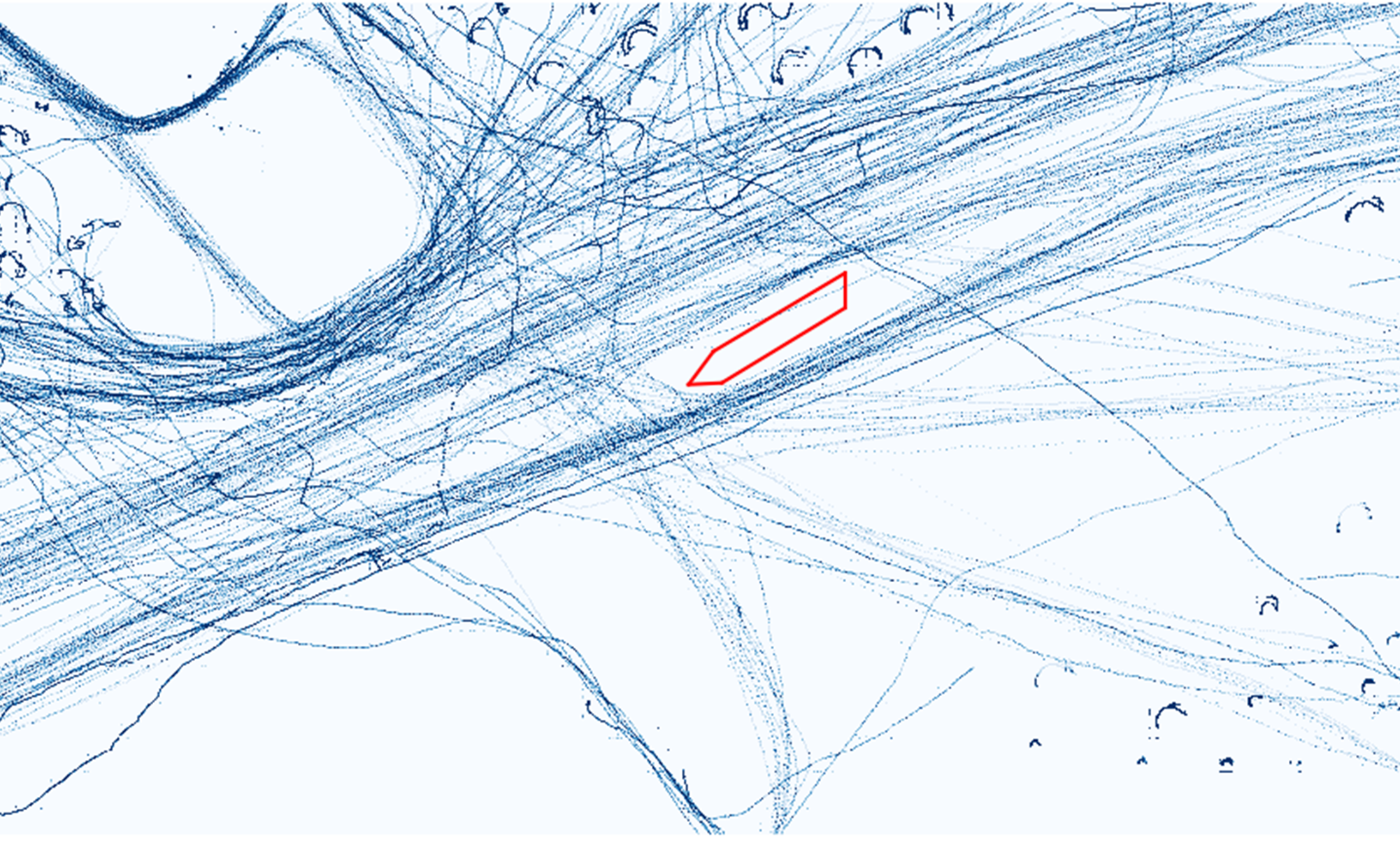}}
~
\subfigure[Vessel Trajectories in August 2017]{
	\label{fig:example:ref}
	\includegraphics[width=0.47\textwidth]{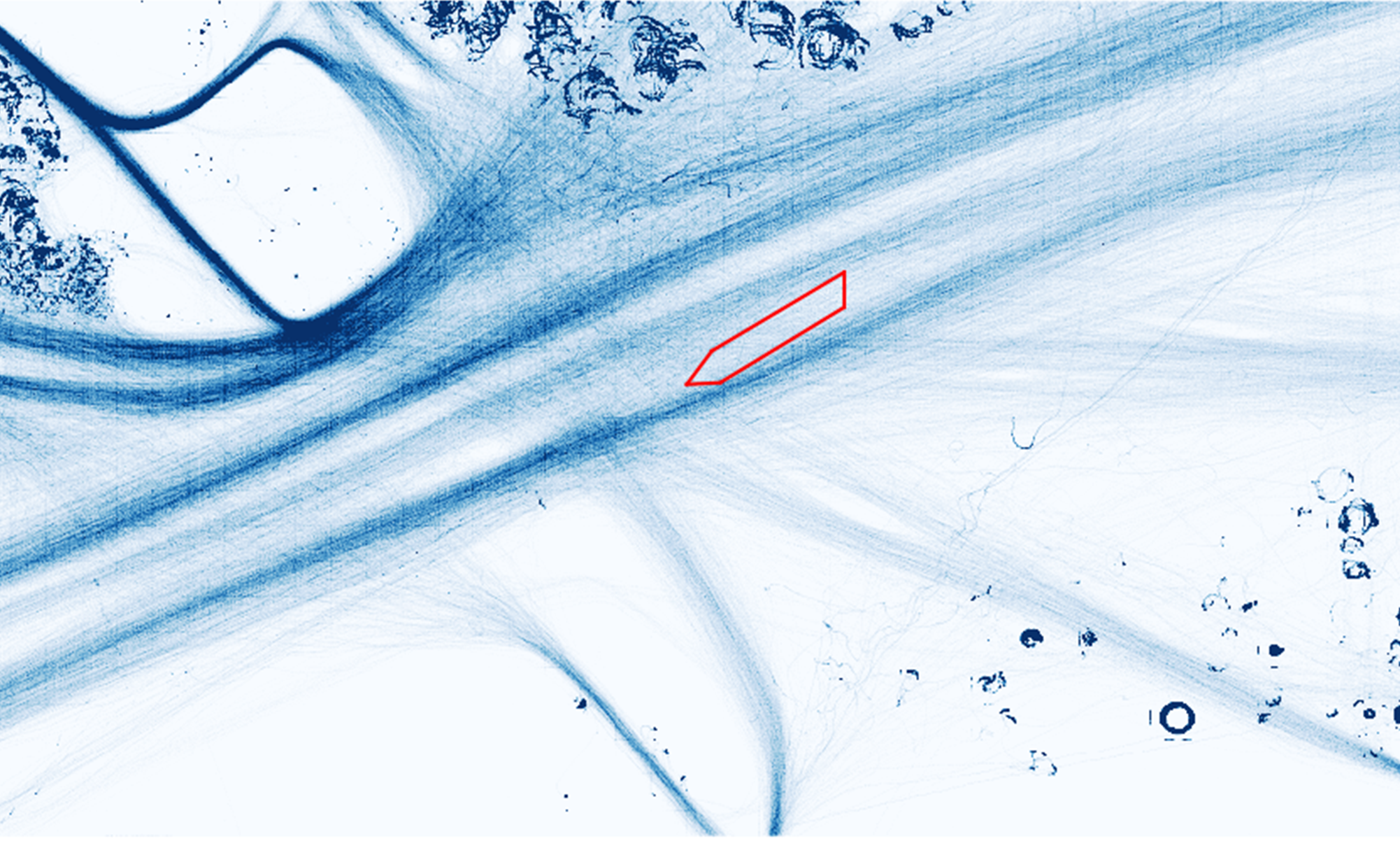}}
\caption{An example of obstacle}
\label{fig:example}
\end{figure}

\begin{example}
\label{example:vessel}
Obstacles are ubiquitous. Figure \ref{fig:example} shows a real-life example of the obstacle. We plot the vessel trajectories in May 2017 and August 2017 in Figures \ref{fig:example:query} and \ref{fig:example:ref}, respectively. According to the official document of Singaporean Notices to Mariners in June 2017,\footnote{\url{https://www.mpa.gov.sg/web/wcm/connect/www/b10f0a7b-09fe-4642-bc30-0282ff8b48f4/notmarijun17.pdf?MOD=AJPERES}.} there is a temporary exclusion zone for operations on a sunken vessel \emph{Thorco Cloud} from March to June 2017, and a red polygon shows its geo-location. From Figure \ref{fig:example}, most trajectories in May 2017 bypass the red polygon, whereas the trajectories in August 2017 can go through this zone as usual. Thus, this zone can be regarded as an obstacle.
\end{example}

According to Example \ref{example:vessel}, we summarize three properties that can quantify an obstacle: (1) \emph{relativity}: an obstacle is a relative concept when comparing two kinds of trajectories; (2) \emph{significance}: the density of different kinds of trajectories significantly deviates on the region; (3) \emph{support}: there should be sufficient observations to support this inference. 
For ease of illustration, we use $\mathcal{T}$ to denote the usual reference trajectories and $\mathcal{Q}$ to indicate the query trajectories. For different query trajectories $\mathcal{Q}$, we can detect different implicit obstacles based on the density variations of $\mathcal{Q}$ compared with $\mathcal{T}$.

\paragraph{Applications}
Obstacle detection arises naturally in many real-life scenarios, such as urban planning and transportation analysis.

\subparagraph{Scenario 1: Urban Planning}
The government can leverage the trajectories from different kinds of anonymous people to detect obstacles for urban planning. For example, suppose there are two kinds of trajectories (e.g., youngsters' trajectories $\mathcal{T}$ and elderlies' trajectories $\mathcal{Q}$) passing through a housing estate. The government can detect implicit obstacles (e.g., steep slope with stairs) for elderlies based on the density variations of $\mathcal{Q}$ compared with $\mathcal{T}$ and redesign the housing estates to make elderly easier to move through. 

\subparagraph{Scenario 2: Transportation Analysis}
Suppose there is a highway road with two lanes. $\mathcal{T}$ denotes a set of car trajectories from suburb to downtown; $\mathcal{Q}$ is a set of car trajectories vice versa. In the morning, the lane of $\mathcal{Q}$ is an obstacle because most people live in the suburbs, and they need to drive to the downtown to work. Thus, the lane of $\mathcal{T}$ has much higher density than that of $\mathcal{Q}$. Similarly, the lane of $\mathcal{T}$ can be detected as an obstacle in the afternoon. Based on such inferences, the traffic management department can change this road as a tidal road.

\paragraph{Why Trajectory}
One might wonder why we detect obstacles based on trajectories but not 2D histograms or road networks. 
The 2D histograms satisfy Scenario 1, but they cannot carry sequential and directional information, which does not satisfy Scenario 2 and is not general enough. 
The road networks are suitable for Scenario 2. However, they cannot model the data that move randomly, such as the trajectories of vessels or pedestrians without the constraint of roads, limiting them to extend to broader scenarios such as Scenario 1. 
Moreover, Compared to the long trajectories with variable sizes, obstacles are often small regions. Thus, we partition trajectories into fixed-size sub-trajectories and use them as the primary data representation.

\paragraph{Research Gap}
Despite its great usefulness, the problem of obstacle detection is still fresh to be solved. Most related works focus on anomaly detection \cite{lee2008trajectory,zhang2011ibat,banerjee2016mantra,hundman2018detecting}, which in general is to find objects with low density. An obstacle, however, does not necessarily have a lower density in reference trajectories $\mathcal{T}$. Another category is avoidance detection \cite{li2013attraction,lettich2016detecting}, which usually detects pairs (or groups) of trajectories that often avoid each other. Nonetheless, the detected trajectory pairs (or groups) should come from the same time period, while the problem of obstacle detection assume that such pairs (or groups) are from two different kinds of trajectories. Therefore, even though the problem of obstacle detection share some similar nature of existing works, it cannot be solved directly by existing algorithms. 

\subsection{Our Contributions}
We first formalize the definition of obstacle. To characterize the obstacles, we design a new normalized Dynamic Time Warping (nDTW) distance and develop the density functions to estimate the density variation of sub-trajectories and their succeeds. The definition can reflect the three properties of an obstacle.

Based on the obstacle definition, we propose a novel framework \textsf{DIOT} to \underline{D}etect \underline{I}mplicit \underline{O}bstacles from \underline{T}rajectories. 
Given a collection of reference sub-trajectories $P(\mathcal{T})$, for a set of query sub-trajectories $P(\mathcal{Q})$, the insight of \textsf{DIOT} is to recursively identify the reference sub-trajectory $t \in P(\mathcal{T})$ whose density variation in $P(\mathcal{Q})$ is significantly larger than that in $P(\mathcal{T})$. 
To efficiently detect implicit obstacles, we design a Depth-First Search (DFS) method to recursively check the $k$ Nearest Neighbors ($k$NNs) of every $q \in P(\mathcal{Q})$ and their $k$NNs and identify the candidate sub-trajectories, until no further new candidates can be found or all query sub-trajectories have been checked. 
Based on our analysis, the obstacle detection can be done in $O(m\log m + m\log n)$ time and $O(m+n)$ space, where $n = \num{P(\mathcal{T})}$ and $m = \num{P(\mathcal{Q})}$. 
We summarize the primary contributions of this work as follows.
\begin{itemize}
\item We discover a real-life problem of obstacle detection, and it has wide applications in many scenarios, such as urban planning and transportation analysis. 
\item We formalize the definition of the obstacle with a new designed nDTW distance and the customized density functions. 
\item We propose an efficient framework \textsf{DIOT} that utilizes the DFS method to detect the implicit obstacles from sub-trajectories.
\item Extensive experiments on two real-life data sets demonstrate the superior performance of \textsf{DIOT}. 
\end{itemize}

\subsection{Related Works}
\vspace{-0.5em}
\paragraph{Anomaly Detection}
A related topic to obstacle detection is the trajectory anomaly detection, which has attracted extensive studies over the past decades \cite{lee2008trajectory,kong2018lotad,zong2018deep,hundman2018detecting,su2019robust}.
Existing techniques for trajectory anomaly detection can be broadly divided into two categories: (1) near-neighbor based and (2) model-based methods. 
Near-neighbor based methods \cite{lee2008trajectory,kong2018lotad} usually use the local neighbors of a trajectory/sub-trajectory to determine whether it is an anomaly.
Recently, due to the prevalence of neural network, numerous model-based methods \cite{hundman2018detecting,zong2018deep,su2019robust} have been proposed for spatial and temporal anomaly detection. Many network structures such as LSTM \cite{hundman2018detecting}, Deep AutoEncoder \cite{zong2018deep}, and Recurrent Neural Network \cite{su2019robust} have been used to detect anomaly.

The obstacle can be seen as a special case of anomaly, or in particular anomalous micro-clusters, where a specific set of trajectories behave differently than the majority of the trajectories. 
However, compared with the anomaly, the concept of an obstacle is based on two kinds of trajectories (relativity). The obstacle does not have a low density in reference trajectories. 
Thus, vanilla anomaly detection methods cannot detect the implicit obstacles effectively.

\paragraph{Avoidance Detection}
Another relevant topic to obstacle detection is the trajectory pattern mining \cite{li2013attraction,lettich2016detecting,he2018detecting,chen2019real}. A notable among them is the avoidance detection \cite{li2013attraction,lettich2016detecting}, which aims to detect pairs (or groups) of trajectories that tend to avoid each other. 
The detected avoidance pattern is usually for a specific time stamp. In contrast, obstacle detection has no such constraint. For example, as shown in Figure \ref{fig:example}, the reference trajectories are from August 2017, while the query trajectories are from May 2017. 

Moreover, avoidance detection returns pairs (or groups) of trajectories, while obstacle detection aims to find a region with a set of reference sub-trajectories. If we use avoidance to detect obstacles, those sub-trajectories that cause obstacles have to be tracked, which is usually not true in real applications. Furthermore, multiple avoidance patterns need to be combined to form an obstacle.
Thus, even though they share a similar nature, avoidance detection methods cannot directly deal with obstacle detection. 

%\subsection{Organization}
%The rest of this work are organized as follows. We first formalize the problem of obstacle detection in Section \ref{sect:problem}. Section \ref{sect:framework} introduces the \textsf{DIOT} framework. Experimental results are reported and analyzed in Section \ref{sect:expt}. Finally, we conclude our work in Section \ref{sect:conclusions}.

%%%%%%%%%%%%%%%%%%%%%%%%%%%%%%%%%%%%%%%%%%%%%%%%%%%%%%%%%%%%%%%%%%%%%%%
%% Problem Definition
%%%%%%%%%%%%%%%%%%%%%%%%%%%%%%%%%%%%%%%%%%%%%%%%%%%%%%%%%%%%%%%%%%%%%%%
\section{Problem Formulation}
\label{sect:problem}
In this section, we formalize the problem of obstacle detection. We first describe some basic concepts about trajectory and sub-trajectory in Section \ref{sect:problem:basic}.  Then, we propose the definitions of the distance function and density function tailored for the sub-trajectories in Sections \ref{sect:problem:distance} and \ref{sect:problem:density}, respectively. Finally, we define the obstacle and the problem of obstacle detection in Section \ref{sect:problem:obstacle}. 

\subsection{Basic Definitions}
\label{sect:problem:basic}
\vspace{-0.5em}
\paragraph{Trajectory and Sub-Trajectory}
A trajectory $T$ is a sequence of points $(t_{(1)}$, $\cdots,t_{(l)})$, where each point $t_{(i)}$ is a $d$-dimensional vector; $l$ is the length of $T$. A sub-trajectory is defined as a consecutive sub-sequence of a trajectory, i.e., $t=T[i:j]$ is a sub-trajectory of $T$ from point $t_{(i)}$ to $t_{(j)}$ such that $1 \leq i < j \leq l$. 

Reference and query trajectories can be considered as two kinds of trajectories from different sources, such as different time periods, different objects, etc. In this paper, we suppose $\mathcal{T}$ and $\mathcal{Q}$ be a collection of reference and query trajectories, respectively. We also assume each point $t_{(i)}$ represents as a $2$-dimensional geo-location (latitude, longitude). \textsf{DIOT} can be easily extended to support obstacle detection for $d$-dimensional points for any $d \geq 3$. 

\paragraph{Partition and Succeed}
Trajectories are often very long and their lengths are variable, but obstacles are usually represented as small regions. It might be hard to utilize the density of long trajectories to detect small implicit obstacles. Thus, before obstacle detection, we define a \emph{partition} operation to split the long trajectories into a set of short sub-trajectories with a fixed size. 
\begin{definition}[Partition]
\label{def:partition}
Given a sliding window $w$ and a step $s$~$(s<w)$, we \emph{partition} a trajectory $T$ into a set of sub-trajectories, i.e., $P(T) = \{t_1,t_2,t_3,\cdots\}$, where $t_1=T[1:w]$, $t_2=T[1+s:w+s]$, $t_3=T[1+2s:w+2s]$, etc. 
\end{definition}

Notice that sometimes $(l-w) \not\equiv 0 \bmod s $. For this case, since $s < w \ll l$, we remove the last few points of $T$ to make it divisible. Thus, each $t_i$ has the same length $w$. For ease of illustration, we use upper case characters (e.g., $T$ and $Q$) to denote trajectories and lower case characters (e.g., $t$ and $q$) to denote sub-trajectories. We use $t$ and $t_i$ interchangeably if no ambiguity.
Furthermore, we define a notion of \emph{succeed} to represent their order relationships.
\begin{definition}[Succeed]
\label{def:succeed}
Given a collection of sub-trajectories $\{t_1,t_2$, $t_3,\cdots\}$ which are partitioned by sequential order from the same trajectory $T$, $t_2$ is the succeed of $t_1$ and $t_3$ is the succeed of $t_2$, i.e., $t_2 = succ(t_1)$ and $t_3=succ(t_2)$.
\end{definition}

After partitioning $\mathcal{T}$ and $\mathcal{Q}$, we get two kinds of sub-trajectories, i.e., $P(\mathcal{T}) = \{P(T) \mid T \in \mathcal{T}\}$ and $P(\mathcal{Q}) = \{P(Q) \mid Q \in \mathcal{Q}\}$. Hereafter, we detect implicit obstacles based on the density variations of $P(\mathcal{Q})$ compared with $P(\mathcal{T})$.

\subsection{Distance Function} 
\label{sect:problem:distance}
As is well known, Dynamic Time Warping (DTW) \cite{myers1981comparative} is one of the most robust and widely used distance functions for the trajectory and time-series data \cite{berndt1994using,ding2008querying,wang2013experimental}. Formally,
\begin{definition}[DTW]
\label{def:dtw}
Given any two sub-trajectories $t=(t_{(1)},\cdots,t_{(l)})$ and $q = (q_{(1)},\cdots,q_{(h)})$, suppose $\norm{t_{(l)}-q_{(h)}}$ is the Euclidean distance between any two points $t_{(l)}$ and $q_{(h)}$. Let $t^{l-1}$ be the first $(l-1)$ points of $t$. The $DTW(t,q)$ is computed as follows:
\begin{displaymath}
DTW(t,q) =
\begin{cases}
\sum_{i=1}^l \norm{t_{(i)}-q_{(1)}}, & \text{if }h=1 \\
\sum_{j=1}^h \norm{t_{(1)}-q_{(j)}}, & \text{if }l=1 \\
\norm{t_{(l)}-q_{(h)}}+\min\{DTW(t^{l-1},q^{h-1}),\\ DTW(t^{l-1},q),DTW(t,q^{h-1})\}. & \text{otherwise}
\end{cases}
\end{displaymath}
\end{definition}

\begin{figure}[h]
\centering
\includegraphics[width=1.0\textwidth]{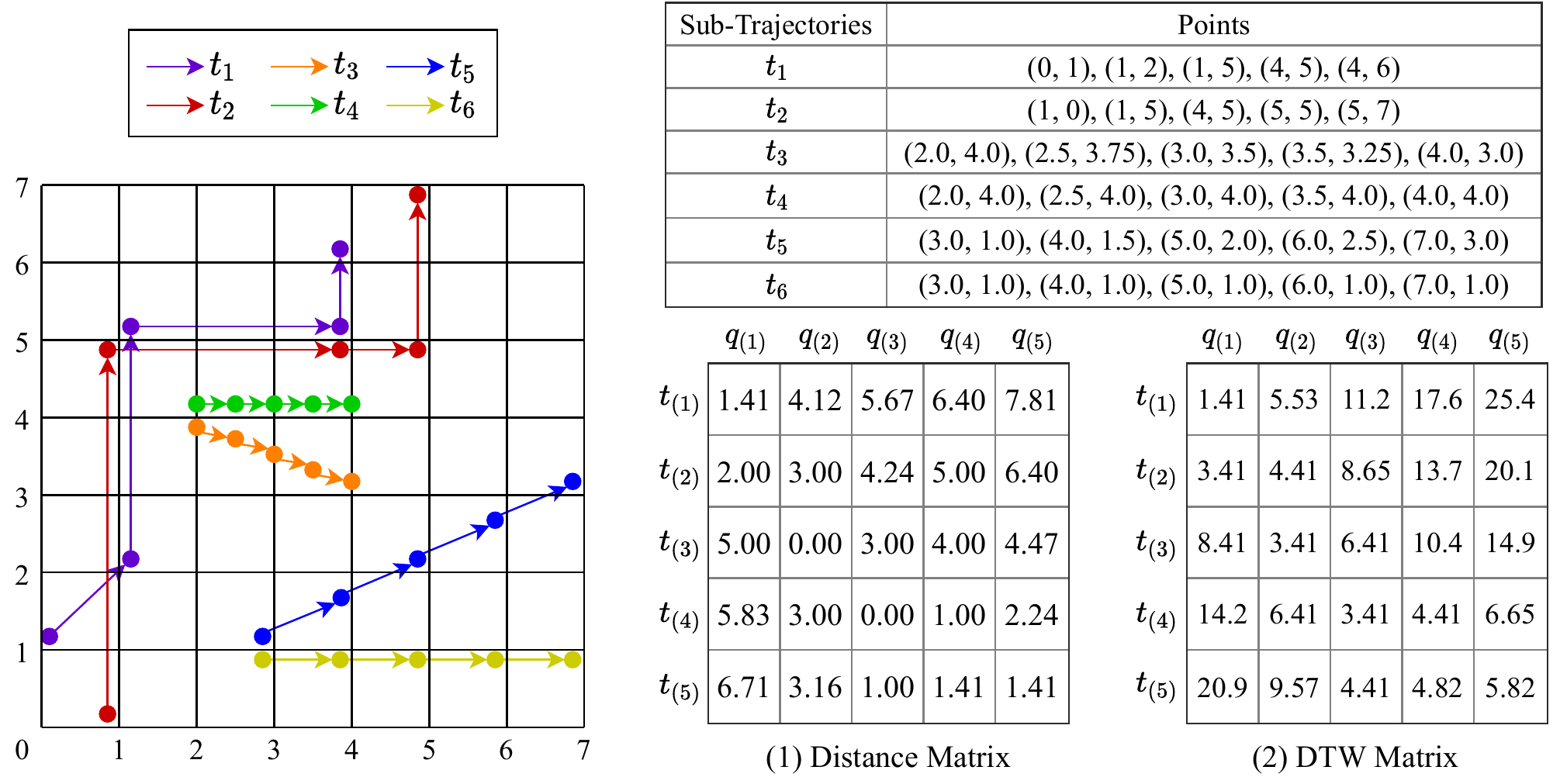}
\caption{An example of sub-trajectories and the computation of DTW}
\label{fig:dtw}
\end{figure}

Given any two sub-trajectories $t$ and $q$ of the same length $w$, we can compute $DTW(t,q)$ in $O(w^2)$ time via dynamic programming. Figure \ref{fig:dtw} shows an example of sub-trajectories and the computation of DTW. Suppose $t=t_1$ and $q=t_2$. Based on Definition \ref{def:dtw}, we compute the point-to-point Euclidean distances and the DTW of $t^i$ and $q^j$ and store those values in a distance matrix and a DTW matrix, respectively. According to the DTW matrix, $DTW(t,q)=5.82$. 

Nonetheless, it might not be sufficient to use DTW as the distance function of sub-trajectories for density estimation. 
For example, as depicted in Figure \ref{fig:dtw}, the pair $(t_3,t_4)$ shows the \emph{same pattern} as $(t_5,t_6)$. They should have the same density, but $DTW(t_3,t_4) < DTW(t_5, t_6)$.
Thus, we propose a \emph{normalized DTW (nDTW)} as the distance function of sub-trajectories for density estimation. Compared with DTW, we use the length of sub-trajectories for normalization.
\begin{definition}[nDTW]
\label{def:ndtw}
Given any two sub-trajectories $t=(t_{(1)},\cdots,t_{(w)})$ and $q = (q_{(1)},\cdots,q_{(w)})$, the $nDTW(t,q)$ is computed as follows:
\begin{displaymath}
nDTW(t,q) = \frac{DTW(t,q)}{\sqrt{\sum_{i=1}^{w-1} \norm{t_{(i)}-t_{(i+1)}}} \sqrt{\sum_{j=1}^{w-1} \norm{q_{(j)}-q_{(j+1)}}}}.
\end{displaymath}
\end{definition}

According to Definition \ref{def:ndtw}, $nDTW(t_3,t_4) = nDTW(t_5,t_6)$. Next, we will apply $nDTW(t,q)$ to the density function such that the pairs of sub-trajectories with the \emph{same pattern} have the \emph{same density} for obstacle detection.

\subsection{Density Function}
\label{sect:problem:density}
To evaluate the density variation of sub-trajectories, we introduce the density functions for the sub-trajectories and their succeeds in this subsection.

\paragraph{Density of Sub-Trajectories}
Given a set of sub-trajectories $P(\mathcal{T})$, we often assume they follow a certain probability density function (PDF) $f_{P(\mathcal{T})}$. However, this PDF is unknown. 
Fortunately, since \emph{each sub-trajectory $t$ can be represented as a $2w$-dimensional vector}, one can adopt the popular Gaussian Kernel Density Estimation (KDE) \cite{zheng2013quality,phillips2018improved} to estimate the density of $t$.
\begin{definition}
\label{def:density-func}
Given a collection of sub-trajectories $P(\mathcal{T})$, for any sub-trajectory $t$, we estimate the density function $f_{P(\mathcal{T})}(t)$ as $\hat{f}_{P(\mathcal{T})}(t)$, i.e.,
\begin{equation}
\label{eqn:density-func}
\hat{f}_{P(\mathcal{T})}(t)= \frac{1}{|P(\mathcal{T})|} \sum_{t_i \in P(\mathcal{T})} \exp(-\frac{{nDTW(t,t_i)}^2}{2 \sigma^2}),
\end{equation}
where $\sigma$ determines the bandwidth of the Gaussian kernel.
\end{definition}

Based on Equation \ref{eqn:density-func}, for any sub-trajectory $t_i \in P(\mathcal{T})$, the contribution of $t_i$ to $\hat{f}_{P(\mathcal{T})}(t)$ decreases dramatically as $nDTW(t,t_i)$ increases. 
In other words, for those $t_i \in P(\mathcal{T})$ that are far from $t$, their contributions to $\hat{f}_{P(\mathcal{T})}(t)$ can be neglected. 
Thus, to reduce the computational cost, we only consider the \emph{$k$ Nearest Neighbors} ($k$NNs) of $t$, i.e., $N_\mathcal{T}(t)$, to compute $\hat{f}_{P(\mathcal{T})}(t)$. The $N_\mathcal{T}(t)$ can be determined by the $k$-Nearest Neighbor Search ($k$-NNS) of $t$ as follows. 
\begin{definition}[$k$-NNS]
\label{def:kNN}
Given a set of sub-trajectories $P(\mathcal{T})$, the $k$-NNS is to construct a data structure which, for any query $t$, finds $k$ sub-trajectories $t_i^* \in P(\mathcal{T})$ $(1 \leq i \leq k)$ such that $nDTW(t,t_i^*) \leq \min_{t_j \in P(\mathcal{T}) \backslash N_\mathcal{T}(t)} nDTW(t,t_j)$, where $N_\mathcal{T}(t) = \{t_i^*\}_{i=1}^k$ are the $k$NNs of $t$. Ties are broken arbitrarily.
\end{definition}

According to Definition \ref{def:partition}, each trajectory has been partitioned into a set of sub-trajectories. Thus, $\hat{f}_{P(\mathcal{T})}(t)$ maybe inflated if some $t_i^* \in N_\mathcal{T}(t)$ are from the same trajectory. In the worst case, all $t_i^* \in N_\mathcal{T}(t)$ are from the same trajectory, then $\hat{f}_{P(\mathcal{T})}(t)$ might be high but the actual density of $t$ is low.

To remedy this issue, we add an extra condition to the $k$ nearest sub-trajectories of $t$ such that they should come from \emph{distinct} trajectories. 
Let $\tilde{N}_\mathcal{T}(t)$ be the $k$ nearest sub-trajectories of $t$ from $k$ distinct trajectories of $\mathcal{T}$. Compared to Definition \ref{def:density-func}, we use $\hat{f}_{\tilde{N}_\mathcal{T}}(t)$ instead of $\hat{f}_{P(\mathcal{T})}(t)$ to estimate the density of $t$:
\begin{equation}
\label{eqn:density-func-t}
\hat{f}_{\tilde{N}_\mathcal{T}}(t) = \frac{1}{|\mathcal{T}|} \sum_{t_i^* \in \tilde{N}_\mathcal{T}(t)} \exp(-\frac{{nDTW(t,t_i^*)}^2}{2 \sigma^2}).
\end{equation}

\paragraph{Density of Succeed Sub-Trajectories}
To evaluate the density variation of $t$, we need to estimate the density of $succ(t)$, which is computed as follows:
\begin{equation}
\label{eqn:density-func-succ-t}
\hat{f}_{\tilde{N}_\mathcal{T},succ}(t) = \frac{1}{|\mathcal{T}|} \sum_{t_i^* \in \tilde{N}_\mathcal{T}(t)} \exp(-\frac{\Delta_i^2}{2 \sigma^2}),
\end{equation}
where $\Delta_i = \max\{{nDTW(t,t_i^*)},{nDTW(succ(t),succ(t_i^*))}\}$.

Compared to $\hat{f}_{\tilde{N}_\mathcal{T}}(t)$, we only uses the succeeds from the same $\tilde{N}_\mathcal{T}(t)$ to estimate $\hat{f}_{\tilde{N}_\mathcal{T},succ}(t)$ because we aim to evaluate the density variation of $t$, so we only consider the density of $succ(t)$ from the same direction of $t$. 
If we consider the $k$NNs of $succ(t)$ (i.e., $\tilde{N}_\mathcal{T}(succ(t))$) to estimate $\hat{f}_{\tilde{N}_\mathcal{T},succ}(t)$, since such $\tilde{N}_\mathcal{T}(succ(t))$ can come from different directions, we might not capture the density variation of $t$. This operation can also save the computational cost to determine $\tilde{N}_\mathcal{T}(succ(t))$.
Moreover, to evaluate $\hat{f}_{\tilde{N}_\mathcal{T},succ}(t)$ precisely, we use $\Delta_i$ in Equation \ref{eqn:density-func-succ-t} to add penalty to $\hat{f}_{\tilde{N}_\mathcal{T},succ}(t)$ if any $succ(t_i^*)$ is no longer close to $succ(t)$. 

\subsection{Obstacle Detection}
\label{sect:problem:obstacle}
Finally, we formalize the definition of obstacle and the problem of obstacle detection. 
Intuitively, the obstacle is a \emph{relative} concept, which is detected from a subset of reference sub-trajectories $P(\mathcal{T})$ such that given some query sub-trajectories $q \in P(\mathcal{Q})$, each $t \in N_\mathcal{T}(q)$ should satisfy two conditions: (1) the density variation of $t$ in $P(\mathcal{T})$ is \emph{significantly} different from that in $P(\mathcal{Q})$; (2) both $\tilde{N}_\mathcal{T}(t)$ and $\tilde{N}_\mathcal{Q}(t)$ should be close to $t$ to \emph{support} condition (1). 
Based on the above analysis, we follow the Association Rule \cite{agrawal1993mining} and DBSCAN \cite{ester1996density} and adopt the standard $z$-test of hypothesis testing \cite{sprinthall1990basic} to define the obstacle. Formally,
\begin{definition}[Obstacle]
\label{def:obstacle}
Given two thresholds $\tau$ ($\tau > 0$) and $\delta$ ($\delta > 0$), obstacles are detected from two kinds of sub-trajectories $P(\mathcal{T})$ and $P(\mathcal{Q})$ \emph{(Relativity)}. An obstacle is a set of last points from a subset of $P(\mathcal{T})$ such that for a subset of close query sub-trajectories $q \in P(\mathcal{Q})$, each $t \in N_\mathcal{T}(q)$ should satisfy:
\begin{itemize}
\item \emph{(Significance)} The density variation score of $t$ is significant, i.e., 
\begin{equation}
\label{eqn:score}
score(t) = \frac{\hat{p}_1-\hat{p}_2}{\sqrt{\hat{p}(1-\hat{p})(\tfrac{1}{\hat{f}_{\tilde{N}_\mathcal{T}}(t)}-\tfrac{1}{\hat{f}_{\tilde{N}_\mathcal{Q}}(t)})}} > \tau,
\end{equation}
where $\hat{p}_1 = \frac{\hat{f}_{\tilde{N}_\mathcal{T},succ}(t)}{\hat{f}_{\tilde{N}_\mathcal{T}}(t)}$, $\hat{p}_2 = \frac{\hat{f}_{\tilde{N}_\mathcal{Q},succ}(t)}{\hat{f}_{\tilde{N}_\mathcal{Q}}(t)}$, and $\hat{p} = \frac{\hat{f}_{\tilde{N}_\mathcal{T}}(t) \cdot \hat{p}_1 + \hat{f}_{\tilde{N}_\mathcal{Q}}(t) \cdot \hat{p}_2}{ \hat{f}_{\tilde{N}_\mathcal{T}}(t)+\hat{f}_{\tilde{N}_\mathcal{Q}}(t)}$. 
\item \emph{(Support)} $\tilde{N}_\mathcal{T}(t)$ and $\tilde{N}_\mathcal{Q}(t)$ are close to $t$, i.e.,
\begin{equation}
\label{eqn:support}
\hat{f}_{\tilde{N}_\mathcal{T}}(t) > \delta\ and\ \hat{f}_{\tilde{N}_\mathcal{Q}}(t) > \delta.
\end{equation}
\end{itemize}
\end{definition}

% explain why we use Equations 4 and to measure significance and support
In Definition \ref{def:obstacle}, inspired by Association Rule, we use the ratio $\hat{p}_1$ and $\hat{p}_2$ to denote the \emph{density variation} of $t$ in $P(\mathcal{T})$ and $P(\mathcal{Q})$, respectively.
We then use the one-sided $z$-test to compute the significance of the density variation (Inequality \ref{eqn:score}). We adopt Inequality \ref{eqn:support} to keep the closeness between $t$ and its nearest sub-trajectories. Additionally, we follow the definition of DBSCAN such that: (1) The query sub-trajectories are close to each other; otherwise, the obstacle can be divided into multiple regions. In the DOIT framework which will be described later, for each query sub-trajectory $q \in P(\mathcal{Q})$, we consider the $k$NNs $N_\mathcal{Q}(q)$ as its close query sub-trajectories. (2) We use the last points of the selected $t \in P(\mathcal{T})$ to construct an obstacle so that it can be of arbitrary shape. 

\begin{figure}[h]
\centering
\subfigure[Obstacle]{%
	\label{fig:property:a}%
	\includegraphics[width=0.23\textwidth]{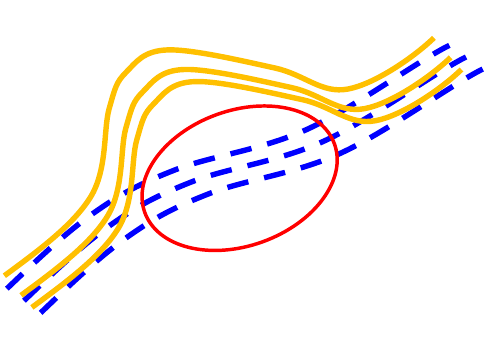}}%
\subfigure[Relativity]{%
	\label{fig:property:b}%
	\includegraphics[width=0.23\textwidth]{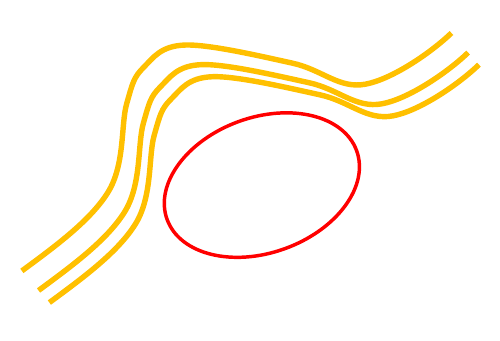}}%
\subfigure[Significance]{%
	\label{fig:property:c}%
	\includegraphics[width=0.23\textwidth]{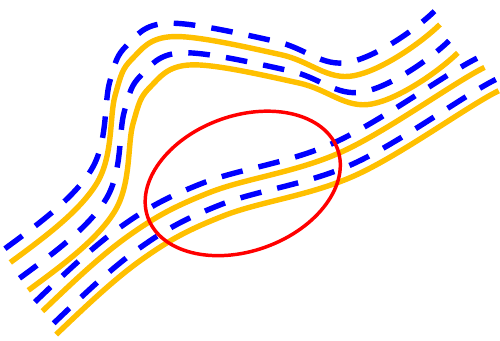}}%
\subfigure[Support]{%
	\label{fig:property:d}%
	\includegraphics[width=0.23\textwidth]{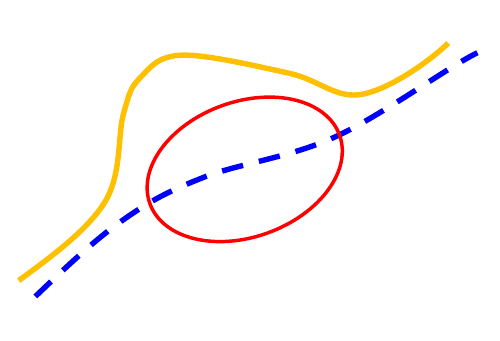}}%
\caption{An illustration of three properties of an obstacle}
\label{fig:property}
\end{figure}

\begin{example}
\label{example:obstacle}
We use Figure \ref{fig:property} to show the basic properties of obstacles and illustrate how Definition \ref{def:obstacle} can fulfill those properties. The blue dash lines and orange solid lines represent the trajectories of $\mathcal{T}$ and $\mathcal{Q}$, respectively. 

From Figure \ref{fig:property:a}, each $t \in P(\mathcal{T})$ passes through the red circle region, while all $q \in P(\mathcal{Q})$ need to bypass this region. Thus, we infer the red circle region is an obstacle. 
As shown in Figure \ref{fig:property:b}, this inference only makes sense under two kinds of sub-trajectories $P(\mathcal{T})$ and $P(\mathcal{Q})$. 
Moreover, Figure \ref{fig:property:c} depicts that if the number of query sub-trajectories in $P(\mathcal{Q})$ passing through the red circle is not significantly smaller than that of $P(\mathcal{T})$, then the red circle region is not an obstacle because Inequality \ref{eqn:score} is not satisfied. 
Finally, from Figure \ref{fig:property:d}, if only one sub-trajectory from $P(\mathcal{T})$ and one from $P(\mathcal{Q})$ satisfy the case depicted in Figure \ref{fig:property:a}, Inequality \ref{eqn:support} fails and the circle is also not an obstacle because the support is not enough.
\hfill $\triangle$ \par 
\end{example}

Since Definition \ref{def:obstacle} is based on two kinds of sub-trajectories, the obstacles are usually different depending on different sets of query sub-trajectories. Thus, we formalize the obstacle detection as an online query problem:
\begin{definition}[Obstacle Detection]
\label{def:obstacle-detection}
Given a set of reference sub-trajectories $P(\mathcal{T})$ and two thresholds $\tau$ ($\tau > 0$) and $\delta$ ($\delta > 0$), the problem of obstacle detection is to construct a data structure which, for a collection of query sub-trajectories $P(\mathcal{Q})$, finds all implicit obstacles as defined in Definition \ref{def:obstacle}.
\end{definition}

%%%%%%%%%%%%%%%%%%%%%%%%%%%%%%%%%%%%%%%%%%%%%%%%%%%%%%%%%%%%%%%%%%%%%%%
%% Framework
%%%%%%%%%%%%%%%%%%%%%%%%%%%%%%%%%%%%%%%%%%%%%%%%%%%%%%%%%%%%%%%%%%%%%%%
\section{The DIOT Framework}
\label{sect:framework}
In this section, we propose a novel framework \textsf{DIOT} for obstacle detection. We first give an overview of \textsf{DIOT} in Section \ref{sect:framework:overview}. Then, we introduce the pre-processing phase and query phase of \textsf{DIOT} in Sections \ref{sect:framework:indexing} and \ref{sect:framework:query}, respectively. Section \ref{sect:framework:optimization} present some strategies to optimize \textsf{DIOT}. Finally, we analyse the time and space complexities in Section \ref{sect:framework:analysis}.

\subsection{Overview}
\label{sect:framework:overview}
The insight of \textsf{DIOT} is to recursively identify the reference sub-trajectories $t \in P(\mathcal{T})$ whose density variation in $P(\mathcal{Q})$ is significantly larger than that in $P(\mathcal{T})$. 
Specifically, according to Definition \ref{def:obstacle}, we aim to find $t \in P(\mathcal{T})$ such that: (1) the density variation score of $t$ is significant, i.e., $score(t) > \tau$; (2) the density of $t$ is large for support, i.e., $\hat{f}_{\tilde{N}_\mathcal{T}}(t) > \delta$ and $\hat{f}_{\tilde{N}_\mathcal{Q}}(t) > \delta$.

\begin{figure}[t]
\centering
\includegraphics[width=1.0\textwidth]{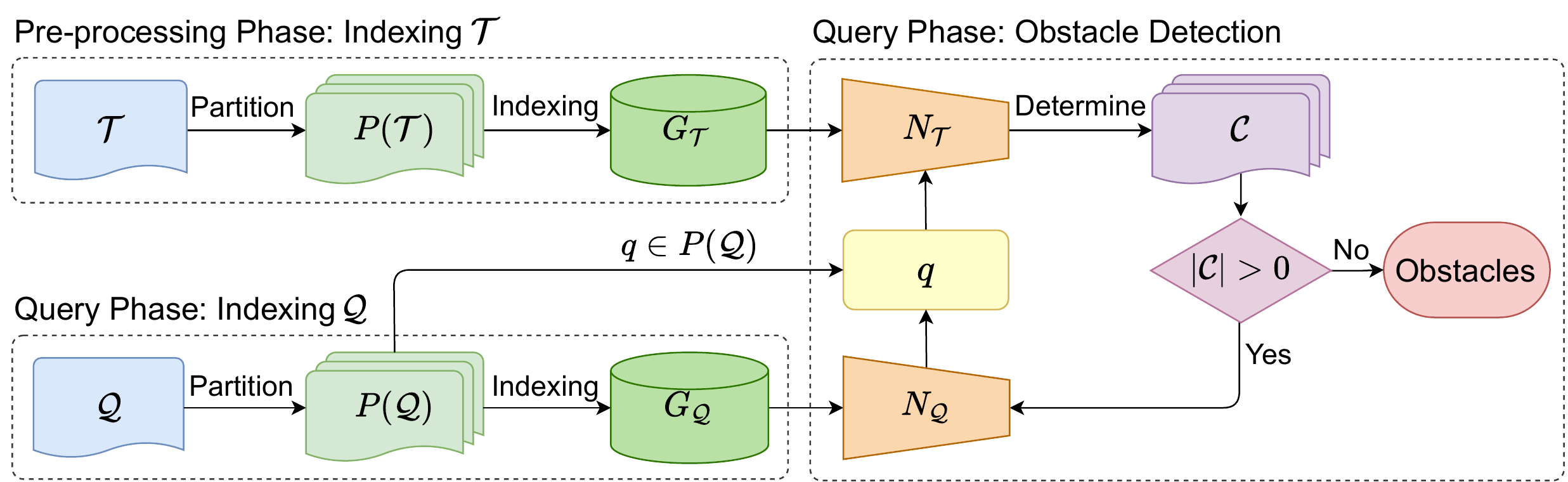}
\caption{An overview of the \textsf{DIOT} framework}
\label{fig:framework}
\end{figure}

\textsf{DIOT} consists of two phases: pre-processing phase and query phase. 
In the pre-processing phase, we partition the reference trajectories $\mathcal{T}$ into a set of sub-trajectories $P(\mathcal{T})$ and build a Hierarchical Navigable Small World (HNSW) graph $G_\mathcal{T}$ \cite{malkov2018efficient} for $P(\mathcal{T})$ so that we can efficiently determine the $k$NNs of sub-trajectories and estimate their densities. 
The query phase also has two steps: indexing $\mathcal{Q}$ and obstacle detection. We use the Depth-First Search (DFS) to determine the candidate sub-trajectories $\mathcal{C}$, that is, recursively check the $k$NNs $N_\mathcal{T}(q)$ of each $q \in P(\mathcal{Q})$ and add $t \in N_\mathcal{T}(q)$ into $\mathcal{C}$ if it satisfies both Inequalities \ref{eqn:score} and \ref{eqn:support}. 
An overview of \textsf{DIOT} is depicted in Figure \ref{fig:framework}.

\subsection{Pre-processing Phase}
\label{sect:framework:indexing}
Given a collection of reference trajectories $\mathcal{T}$, we first follow Definition \ref{def:partition} and partition $\mathcal{T}$ into a set of sub-trajectories $P(\mathcal{T})$. Then, we build an HNSW graph $G_{\mathcal{T}}$ for $P(\mathcal{T})$. The pseudo-code of indexing $\mathcal{T}$ is shown in Algorithm \ref{alg:indexing}. 

We choose HNSW \cite{malkov2018efficient} under two considerations: (1) HNSW is a $k$-Nearest Neighbor Graph ($k$-NNG) based method \cite{dong2011efficient,malkov2018efficient,fu2019fast,zhao2020song}, which is very efficient for $k$-NNS \cite{aumuller2020ann}. (2) Compared with the methods based on Locality-Sensitive Hashing (LSH) \cite{indyk1998approximate,datar2004locality,gan2012locality,sun2014srs,huang2015query,huang2017query,lei2019sublinear,lu2020vhp,zheng2020pm,lei2020locality,huang2021point} and Product Quantization \cite{jegou2011product,johnson2019billion,wang2020deltapq}, the HNSW graph $G_\mathcal{T}$ directly stores the $k$NNs $N_\mathcal{T}(t)$ of $t$. Thus, we can directly retrieve the $N_\mathcal{T}(t)$ for all $t\in P(\mathcal{T})$ without conducting $k$-NNS again in the query phase. 
Notice that HNSW uses a priority queue to perform $k$-NNS. To determine $\tilde{N}_\mathcal{T}(t)$, we add one more condition to the priority queue to check whether the new candidate comes from the same trajectory of the old ones. 

\begin{algorithm}[t]
\caption{Indexing($\mathcal{T},w,s,k$)}
\label{alg:indexing}
$P(\mathcal{T})=\emptyset$\;
\ForEach{$t \in \mathcal{T}$}{
	$P(t) = Partition(t, w, s)$\;
	$P(\mathcal{T}) = P(\mathcal{T}) \cup P(t)$\;
}
Build an HNSW graph $G_\mathcal{T}$ for $P(\mathcal{T})$\;
\Return $P(\mathcal{T})$ and $G_\mathcal{T}$\;
\end{algorithm}

\subsection{Query Phase}
\label{sect:framework:query}
\vspace{-0.5em}
\paragraph{Indexing $\mathcal{Q}$}
Given a set of query trajectories $\mathcal{Q}$, we partition $\mathcal{Q}$ into a collection of query sub-trajectories $P(\mathcal{Q})$ and build an HNSW graph $G_\mathcal{Q}$ for $P(\mathcal{Q})$. The operations are similar to Algorithm \ref{alg:indexing}.
Notice that $|P(\mathcal{Q})|$ cannot be neglected. If without $G_\mathcal{Q}$, we might require $O(|P(\mathcal{Q})|)$ time to determine $\tilde{N}_\mathcal{Q}(t)$ (with the brute-force method). Thus, even though in the query phase, building $G_\mathcal{Q}$ for $P(\mathcal{Q})$ is still beneficial to reduce the computational cost. 

\paragraph{Obstacle Detection}
After indexing $\mathcal{Q}$, we initialize an empty set $\mathcal{S}$ to store obstacles and use a bitmap to flag each query sub-trajectory $q \in P(\mathcal{Q})$ checked or not. 
To detect the implicit obstacles, we find the candidate sub-trajectories $\mathcal{C}$ using the DFS method for each $q \in P(\mathcal{Q})$; then, we construct an obstacle $O$ by the last points of $\mathcal{C}$ and add $O$ to $\mathcal{S}$ if $\num{\mathcal{C}} > 0$. We return $\mathcal{S}$ as the answer. The pseudo-code of obstacle detection is shown in Algorithm \ref{alg:obstacle_detection}. 

\subparagraph{Depth-First Search (DFS)}
Next, we illustrate how to find candidates with the DFS method. 
Given a query sub-trajectory $q \in P(\mathcal{Q})$, we first conduct $k$-NNS and determine its $k$NNs $N_\mathcal{T}(q)$ from $G_\mathcal{T}$. 
For each $t \in N_\mathcal{T}(q)$, we compute $\hat{f}_{\tilde{N}_\mathcal{T}}(t)$, $\hat{f}_{\tilde{N}_\mathcal{T},succ}(t)$, $\hat{f}_{\tilde{N}_\mathcal{Q}}(t)$, $\hat{f}_{\tilde{N}_\mathcal{Q},succ}(t)$ and validate whether both Inequalities \ref{eqn:score} and \ref{eqn:support} are satisfied or not. We add $t$ to $\mathcal{C}$ if both of them are satisfied. 
After checking all $t \in N_\mathcal{T}(q)$, if $\mathcal{C}$ is not empty, i.e., $\num{\mathcal{C}} > 0$, which means there may exist an obstacle, we continue to find the candidate sub-trajectories from the \emph{close query sub-trajectories} of $q$, i.e., its $k$NNs $N_\mathcal{Q}(q)$, until no further new candidate can be found or all $N_\mathcal{Q}(q)$'s have been checked. The pseudo-code of the DFS to determine $\mathcal{C}$ is depicted in Algorithm \ref{alg:find_candidates}.

\begin{algorithm}[tb]
\caption{ObstacleDetection($P(\mathcal{T}),P(\mathcal{Q}),G_\mathcal{T},G_\mathcal{Q},\tau,\delta$)}
\label{alg:obstacle_detection}
$\mathcal{S} = \emptyset$\;
Set $flag[q] =$ \emph{false} for each $q \in P(\mathcal{Q})$ \Comment*[r]{bitmap of $P(\mathcal{Q})$}
\ForEach{$q \in P(\mathcal{Q})$}{
	\If{$!flag[q]$}{
		$\mathcal{C}$ $\leftarrow$ FindCandidates($q,flag,G_\mathcal{T},G_\mathcal{Q},\tau,\delta$)\;
		\If{$\num{\mathcal{C}} > 0$}{
			Determine an obstacle $O$ from $\mathcal{C}$\;
			$\mathcal{S} = \mathcal{S} \cup O$\;
		}
	}
}
\Return $\mathcal{S}$\;
\end{algorithm}

\begin{algorithm}[tb]
\caption{FindCandidates($q,flag,G_\mathcal{T},G_\mathcal{Q},\tau,\delta$)}
\label{alg:find_candidates}
$\mathcal{C}=\emptyset$;~$flag[q] = true$\;
Find $N_\mathcal{T}(q)$ from $G_\mathcal{T}$\;
\ForEach{$t \in N_\mathcal{T}(q)$}{
	Determine $\tilde{N}_\mathcal{T}(t)$ from $G_\mathcal{T}$ and $\tilde{N}_\mathcal{Q}(t)$ from $G_\mathcal{Q}$\;
	\If{$score(t) > \tau$~\textbf{and}~$\hat{f}_{\tilde{N}_\mathcal{T}}(t) > \delta$~\textbf{and}~$\hat{f}_{\tilde{N}_\mathcal{Q}}(t) > \delta$}{
		$\mathcal{C} = \mathcal{C} \cup t$\;
	}
}
\If{$\num{\mathcal{C}} > 0$}{
	find $N_\mathcal{Q}(q)$ from $G_\mathcal{Q}$ \Comment*[r]{Find the close query sub-trajectories of $q$}
	\ForEach{$q_i \in N_\mathcal{Q}(q)$} {
		\If {$!flag[q_i]$}{
			$\mathcal{C}_i \gets$ FindCandidates($q_i,flag,G_\mathcal{T},G_\mathcal{Q},\tau,\delta$)\;
			$\mathcal{C} = \mathcal{C} \cup \mathcal{C}_i$\;
		}
	}
}
\Return $\mathcal{C}$\;
\end{algorithm}

\subparagraph{Remarks}
To find the close query sub-trajectories, one may also consider the Range Neighbor Search (RNS) with a pre-specified distance threshold. 
We choose $k$-NNS because it can fix the number of neighbors (i.e., $k$) for the DFS. In contrast, the number of neighbors returned by the RNS is non-trivial to control, which may lead to extra cost to check the unrelated reference sub-trajectories $t \in P(\mathcal{T})$. If two sub-trajectories are close enough, it can be expected that the DFS method can reach most of those returned by the RNS. Thus, compared with RNS, the $k$-NNS is more suitable to the \textsf{DIOT} framework.

\subsection{Optimizations}
\label{sect:framework:optimization}
The basic obstacle detection algorithm can performed well on many data sets. We now develop four insightful strategies for further optimization.

\paragraph{Pre-compute $\tilde{N}_\mathcal{T}(t)$}
Referring to line 4 in Algorithm \ref{alg:find_candidates}, we need to conduct distinct $k$-NNS twice for each reference sub-trajectory $t \in N_\mathcal{T}(q)$, i.e., find $\tilde{N}_\mathcal{T}(t)$ from $G_\mathcal{T}$ and $\tilde{N}_\mathcal{Q}(t)$ from $G_\mathcal{Q}$, respectively. Notice that the operation to find the $k$ distinct reference sub-trajectories $\tilde{N}_\mathcal{T}(t)$ from $G_\mathcal{T}$ is independent of $\mathcal{Q}$. Thus, we can determine $\tilde{N}_\mathcal{T}(t)$ for each $t \in P(\mathcal{T})$ in the pre-processing phase. Although the query time complexity remains the same, this strategy can save a large amount of running time as it is a very frequent operation.

\paragraph{Build a bitmap of $P(\mathcal{T})$} 
In Algorithms \ref{alg:obstacle_detection} and \ref{alg:find_candidates}, some reference sub-trajectories $t \in P(\mathcal{T})$ might be checked multiple times. For example, suppose the query sub-trajectories $q_1,q_2 \in P(\mathcal{Q})$ and they are close to each other. If $t \in N_\mathcal{T}(q_1)$, it is very likely that $t \in N_\mathcal{T}(q_2)$. As such, we need to check $t$ twice. 
To avoid this case, similar to the operation to $P(\mathcal{Q})$, we also build a bitmap for $P(\mathcal{T})$ in Algorithm \ref{alg:obstacle_detection} to identify whether each $t \in P(\mathcal{T})$ is checked or not. If $t$ has already been checked, we will not check it again to avoid the redundant computations.

\paragraph{Skip the close $q_i \in N_\mathcal{Q}(q)$} 
As we call the DFS method (Algorithm \ref{alg:find_candidates}) recursively, fewer new $t$'s will be added to $\mathcal{C}$. 
Thus, we do not need to consider all $q_i \in N_\mathcal{Q}(q)$ in each recursion as many $t \in N_\mathcal{T}(q_i)$'s have already been checked in previous recursions. 
Thus, before the recursion of $q_i$, we first check its closeness to $q$. Let $\epsilon$ be a small distance threshold. If $nDTW(q_i,q) < \epsilon$, since $N_\mathcal{T}(q_i)$ are almost identical to $N_\mathcal{T}(q)$, we set $flag[q_i]=true$ and skip this recursion. 

\paragraph{Skip the close $t \in N_\mathcal{T}(q)$} Similar to the motivation of skipping the close query sub-trajectories $q_i \in N_\mathcal{Q}(q)$, we do not need to check all $t \in N_\mathcal{T}(q)$. 
Specifically, for each $t \in N_\mathcal{T}(q)$, we first consider its $k$NNs $N_\mathcal{T}(t)$: if there exists a $t_i \in N_\mathcal{T}(t)$ that has been checked and $nDTW(t,t_i) < \epsilon$, we follow the same operation of $t_i$ to $t$ to avoid the distinct $k$-NNS and the validation of Inequalities \ref{eqn:score} and \ref{eqn:support}.

\subsection{Complexity Analysis}
\label{sect:framework:analysis}
\vspace{-0.5em}
\paragraph{Time Complexity}
According to HNSW \cite{malkov2018efficient}, for the low-dimensional data sets with $n$ objects, the graph construction and $k$-NNS requires $O(n \log n)$ and $O(\log n)$ time, respectively. 
Let $n = \num{P(\mathcal{T})}$ and $m = \num{P(\mathcal{Q})}$. 
Each sub-trajectory $t \in P(\mathcal{T})$ and $q \in P(\mathcal{Q})$ is a $2w$-dimensional vector. We set $w=6$ in the experiments. Thus, our analysis satisfies the low-dimensional assumption.

In the pre-processing phase, we need $O(nw)$ time to partition $\mathcal{T}$ into $P(\mathcal{T})$ and $O(n\log n)$ time to construct $G_\mathcal{T}$. 
Moreover, as we only add one more condition in priority queue for filtering, the time complexity of $k$-NNS to determine $\tilde{N}_\mathcal{T}(t)$ is also $O(\log n)$. Thus, we take $O(n\log n)$ time to determine $\tilde{N}_\mathcal{T}(t)$ for all reference sub-trajectories $t\in P(\mathcal{T})$. 
Since $w = O(1)$, the indexing time complexity is $O(n\log n)$. 

In the query phase, we spend $O(m\log m)$ time to index $P(\mathcal{Q})$. In the worst case, we need $O(m\log n)$ time to check \emph{all} $q \in P(\mathcal{Q})$ for obstacle detection as determining $N_\mathcal{T}(q)$ takes $O(\log n)$ time; 
moreover, we need to check $k$ \emph{distinct} sub-trajectories in $N_\mathcal{T}(q)$ for \emph{all} $q \in P(\mathcal{Q})$, which requires $O(km\log m)$ time as finding $\tilde{N}_\mathcal{Q}(t)$ for each $t \in N_\mathcal{T}(q)$ needs $O(\log m)$ time. In practice, we set $k=8$. Since $k = O(1)$, the query time complexity is $O(m\log m + m\log n)$.

\paragraph{Memory Cost}
The memory cost consists of three parts: (1) $O(nw + mw)$ space to store $P(\mathcal{T})$ and $P(\mathcal{Q})$; (2) $O(nk + mk)$ space to store $G_\mathcal{T}$ and $G_\mathcal{Q}$ \cite{malkov2018efficient}; (3) $O(n+m)$ space to store the bitmaps of $P(\mathcal{T})$ and $P(\mathcal{Q})$. Since $w = O(1)$ and $k = O(1)$, the total memory cost is $O(n+m)$.

%%%%%%%%%%%%%%%%%%%%%%%%%%%%%%%%%%%%%%%%%%%%%%%%%%%%%%%%%%%%%%%%%%%%%%%
%% Experiments
%%%%%%%%%%%%%%%%%%%%%%%%%%%%%%%%%%%%%%%%%%%%%%%%%%%%%%%%%%%%%%%%%%%%%%%
\section{Experiments}
\label{sect:expt}
In this section, we conduct experiments on two real-life data sets and study the performance of \textsf{DIOT} for obstacle detection. 

\subsection{Experimental Setup}
\label{sect:expt:setup}
\vspace{-0.5em}
\paragraph{Experiments Environment}
All methods are implemented in C++ and compiled with g++-8 using -O3 optimization. We conduct all experiments in a single thread on a machine with Intel Core i7 CPU and 64 GB memory, running on Ubuntu 18.04.

\paragraph{Evaluation Measures}
We use precision, recall, F1-score, and the query time to evaluate the performance of \textsf{DIOT}. The precision (recall) is computed by the number of matched obstacles over the number of returned obstacles (ground truths). The matched obstacles are the returned obstacles that intersect with the ground truth area, and their directions are towards the ground truths. 
The precision, recall, and F1-score are shown in percentage (\%). The query time refers to the wall-clock time to detect implicit obstacles without indexing $\mathcal{Q}$.

\begin{table}[tb]
\centering
\caption{The statistics of data sets and their indexing time (ITime, in Seconds)}
\label{tab:stat}
\begin{tabular}{ccccccc} \toprule
Data Sets & Query Sets & $\vert P(\mathcal{T}) \vert$ & $\vert P(\mathcal{Q}) \vert$ & \#Obstacles & ITime $P(\mathcal{T})$ & ITime $P(\mathcal{Q})$ \\ \midrule
\multirow{3}{*}{Vessel}
& Harita Berlian & 5,305,225  & 393,054 & 1  & 1522.81 & 99.32 \\
& Thorco Cloud   & 4,991,934  & 41,976  & 1  & 1405.75 & 10.15 \\
& Cai Jun 3      & 4,991,934  & 42,179  & 1  & 1410.28 & 9.15  \\
\multirow{2}{*}{Taxi}
& Morning ERP    & 266,761    & 20,875  & 51 & 69.85   & 4.42  \\
& Afternoon ERP  & 266,761    & 30,057  & 50 & 69.41   & 6.58  \\  \bottomrule 
\end{tabular}
\end{table}

\paragraph{Data Sets}
We use two real-life data sets Vessel and Taxi for validation. In the following, we introduce how to get the reference trajectories, query trajectories, and the ground truths.

\subparagraph{Vessel} 
This data set consists of a collection of GPS records of the vessels near Singapore Strait during May to September 2017. We investigate the temporary and preliminary notices that are related to certain obstacles of Singaporean Notices to Mariners; Then, we find three sunken vessels, i.e., \emph{Harita Berlian, Thorco Cloud, and Cai Jun 3}, with available operating geo-location area. Thus, we select the trajectories in non-operating time as references and the trajectories around the operating region in the operating time as queries. Since vessels would start to avoid the operating region far behind the exact operating region, we enlarge each ground truth region by $2~km$.

\subparagraph{Taxi} 
It is a set of trajectories of 14,579 taxis in Singapore over one week \cite{wu2012taxi}. We study the effect of Electronic Road Pricing (ERP) gantries,\footnote{\url{https://onemotoring.lta.gov.sg/content/onemotoring/home/driving/ERP/ERP.html}} which is an electronic system of road pricing in Singapore. We select the taxi trajectories with free state as references and the trajectories in \emph{morning peak hour} and \emph{afternoon peak hour} when the ERP is working as queries. We suppose taxi drivers do not pass through the ERP gantries when the taxi is free during the ERP operating hours. Thus, we can use the location of working ERP gantries as the ground truths. Since the ERP is a point, we enlarge each ERP to the road segment it belongs to as the ground truth obstacle. 

We use interpolation to align the trajectories with fixed sample rate. Due to the different nature of Vessel and Taxi, we use $600$ and $30$ seconds as the interval of interpolation, respectively. We also set $w = 6$ and $s = 1$ for trajectory partitioning to get the sub-trajectories. Table \ref{tab:stat} summarizes the statistics of data sets. Moreover, we observe that the $k$ value of distinct $k$-NNS is not very sensitive to the results of \textsf{DIOT}. Thus, we simply use $k=8$ for distinct $k$-NNS. 

\begin{table}[t]
\centering
\caption{The results of quantitative analysis, where Prec, Rec, F1, QTime refer to precision, recall, F1-score, and the query time (in Seconds), respectively.}
\label{tab:results}
%\begin{tabular}{*{9}c} \toprule
\begin{tabular}{>{\hfil}p{65pt}<{\hfil} >{\hfil}p{30pt}<{\hfil} >{\hfil}p{30pt}<{\hfil} >{\hfil}p{30pt}<{\hfil} >{\hfil}p{35pt}<{\hfil} >{\hfil}p{30pt}<{\hfil} >{\hfil}p{30pt}<{\hfil} >{\hfil}p{30pt}<{\hfil} >{\hfil}p{35pt}<{\hfil}} \toprule
\multirow{2}{*}{Query Set} & \multicolumn{4}{c}{\textsf{DIOT} without optimization} &\multicolumn{4}{c}{\textsf{DIOT} with optimization}  \\ \cmidrule(lr){2-5} \cmidrule(lr){6-9}
 & Prec  & Rec & F1 & QTime & Prec  & Rec & F1 & QTime \\ \midrule
Harita Berlian& 100.0& 100.0 & 100.0 & 209.16 & 100.0 & 100.0 & 100.0 & 109.87 \\
Thorco Cloud  & 50.0 & 100.0 & 66.7  & 18.46  & 100.0 & 100.0 & 100.0 & 12.31  \\
Cai Jun 3     & 25.0 & 100.0 & 40.0  & 15.06  & 20.0  & 100.0 & 33.3  & 7.23   \\
Morning ERP   & 51.3 & 88.2  & 64.9  & 18.14  & 50.0  & 82.4  & 62.2  & 4.67   \\
Afternoon ERP & 41.7 & 68.0  & 51.7  & 25.47  & 47.6  & 52.0  & 49.7  & 7.32   \\  \bottomrule
\end{tabular}
\end{table}

\subsection{Quantitative Analysis}
\label{sect:expt:quantitative}
We first conduct the quantitative analysis of \textsf{DIOT}. We report the highest F1-scores of \textsf{DIOT} from a set of $\delta\in \{0.5,1.0,\cdots,4.0\}$ and $\tau \in \{1.282, 1.645, 1.960$, $2.326, 2.576\}$ using grid search.\footnote{Specifically, we select $\delta=\{3.5, 1.5, 4.0, 2.0, 2.0\}$ and $\tau=\{1.906, 1.960, 1.960, 1.960$, $1.645\}$ for the five query sets, respectively.} 
The results are depicted in Tables \ref{tab:stat} and \ref{tab:results}.

For Vessel, since each query has only one ground truth obstacle, \textsf{DIOT} can detect all of them with 100\% recall. As the obstacle pattern of Harita Berlian is obvious, its F1-score is uniformly higher than that of Thorco Cloud and Cai Jun 3. \textsf{DIOT} has a relatively lower F1-score for Cai Jun 3 because its operating area is not at the centre of the vessel route. 

For Taxi, the obstacles we found are correlated to the location of ERP gantries. More than 50\% and 40\% returned obstacles fit the location of ERP gantries for Morning and Afternoon ERP queries, respectively. These results validate our assumption that taxi drivers tend to avoid the ERP gantries when their taxis are free.

Table \ref{tab:results} also shows that \textsf{DIOT} with optimization is $2 \sim 4$ times faster than that without optimization under the similar accuracy. This confirms the effect of our proposed optimization strategies. %The analysis of these obstacle patterns and correlations might be useful for city planners and transportation analysis.

\subsection{Case Studies}
Next, we conduct case studies for some typical obstacles found from Vessel and Taxi to validate the actual performance of \textsf{DIOT}.

\begin{figure}[t]
\centering
\subfigure[Case study of Thorco Cloud]{
	\label{fig:case_study:vessel1}
	\includegraphics[width=0.47\textwidth]{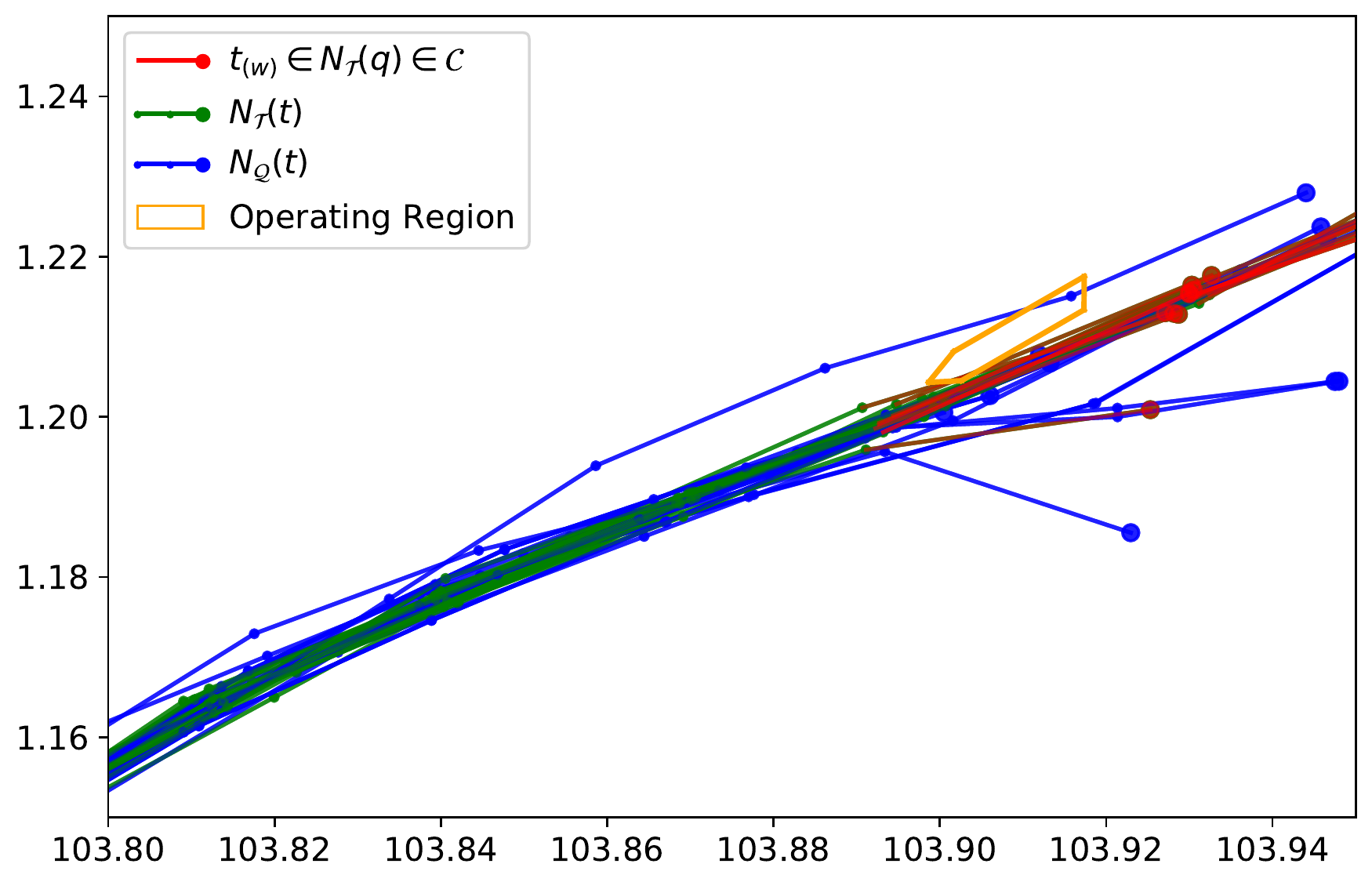}}
~
\subfigure[Case study of Morning ERP]{
	\label{fig:case_study:taxi_morning}
	\includegraphics[width=0.47\textwidth]{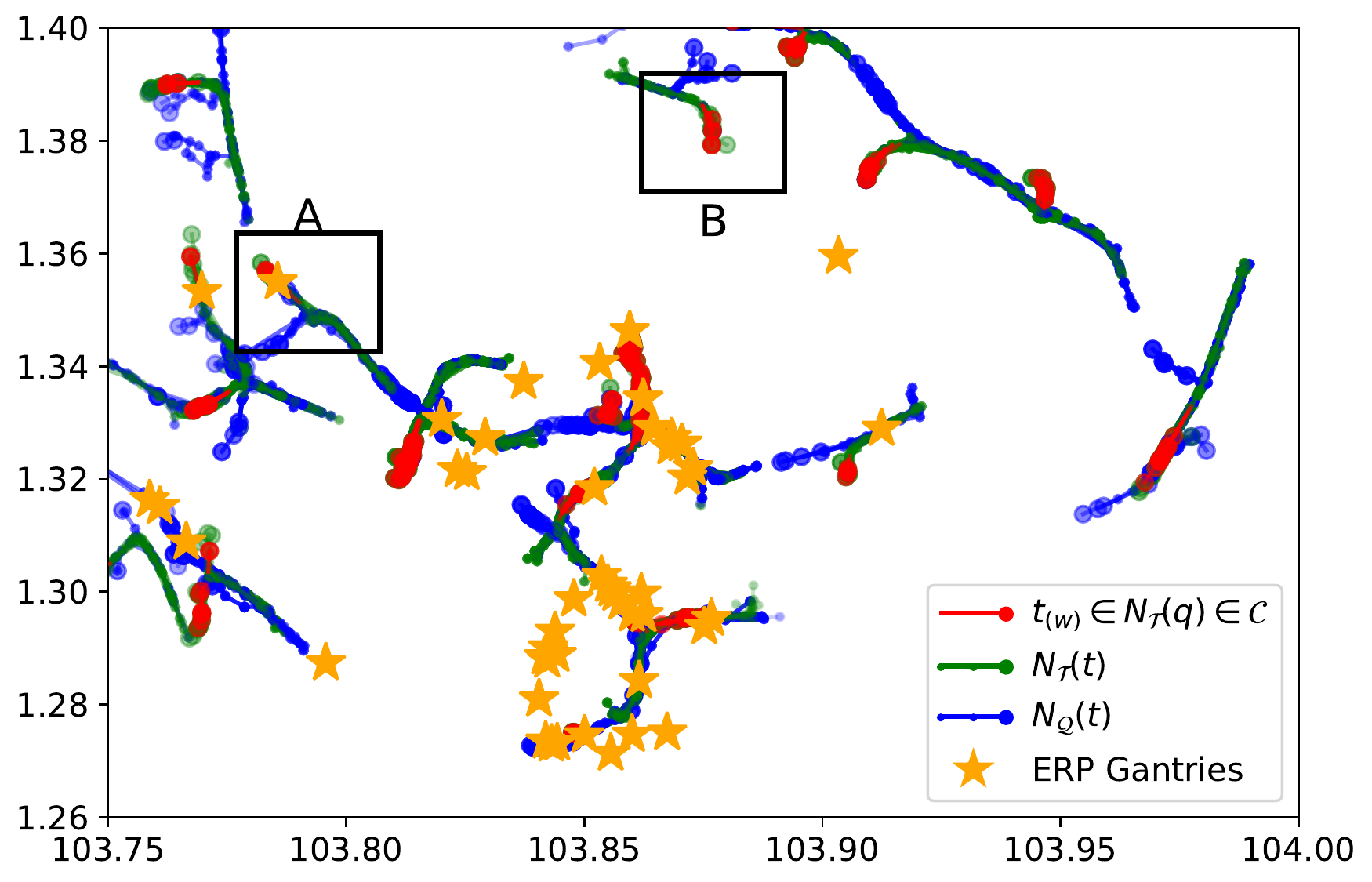}} 
\caption{Case Study}
\label{fig:case_study}
\end{figure}

\paragraph{Vessel: Thorco Cloud}
Figure \ref{fig:case_study:vessel1} shows a real-life obstacle example discussed in Section \ref{sect:intro}. The orange polygon represents the operating area (actual obstacle). The red, green, and blue curves denote the trajectories that respectively present in the returned obstacles, $N_{\mathcal{T}}(t)$, and $N_{\mathcal{Q}}(t)$ of Thorco Cloud; the circles indicate the direction of trajectories. One can regard the convex hull formed by the red circles as the returned obstacle. 
As can be seen, during the operating time, the vessels (blue curves) have clear pattern to avoid the operating area, while during the non-operating time, the vessels (green curves) move freely. This discrepancy is successfully captured by \textsf{DIOT}, and the location of the detected obstacle region (red circles shown in Figure \ref{fig:case_study:vessel1}) fits the operating area. 

\paragraph{Taxi: Morning ERP}
Figure \ref{fig:case_study:taxi_morning} depicts the typical obstacles caused by ERP gantries. The orange stars are the location of ERP gantries. One can find explicit correlations between the returned obstacles and the ERP gantries. 
For example, as shown in Rectangle A, the star represents the ERP gantry in the Bukit Timah Expressway street whose operating time is 7:30--9:00 am weekdays. The query trajectories (blue curves) have significantly less tendency to go towards the ERP gantry. 
Moreover, some obstacles that are not directly associated to the ERP gantries might be caused by the ERP gantry as well. 
For instance, the detected obstacle in Rectangle B is directly towards to the Central Express Street in Singapore that ends with some ERP gantries. Thus, their correlation might be even higher than the precision and recall values shown in Table \ref{tab:results}.

\subsection{Effects of $\delta$ and $\tau$}
\label{sect:expt:para}
Finally, we study the effects of $\delta$ and $\tau$. 
We first fix $\tau=1.645$, which is the critical value for 95\% one-tailed $z$-test and tune $\delta \in \{0.5, 1.0, \cdots, 4.0\}$ to study its effect to the \textsf{DIOT} framework. 

\begin{figure}[t]
\centering
\subfigure[F1-score vs.\ $\delta$]{
	\label{fig:param:f1score_delta}
	\includegraphics[width=0.235\textwidth]{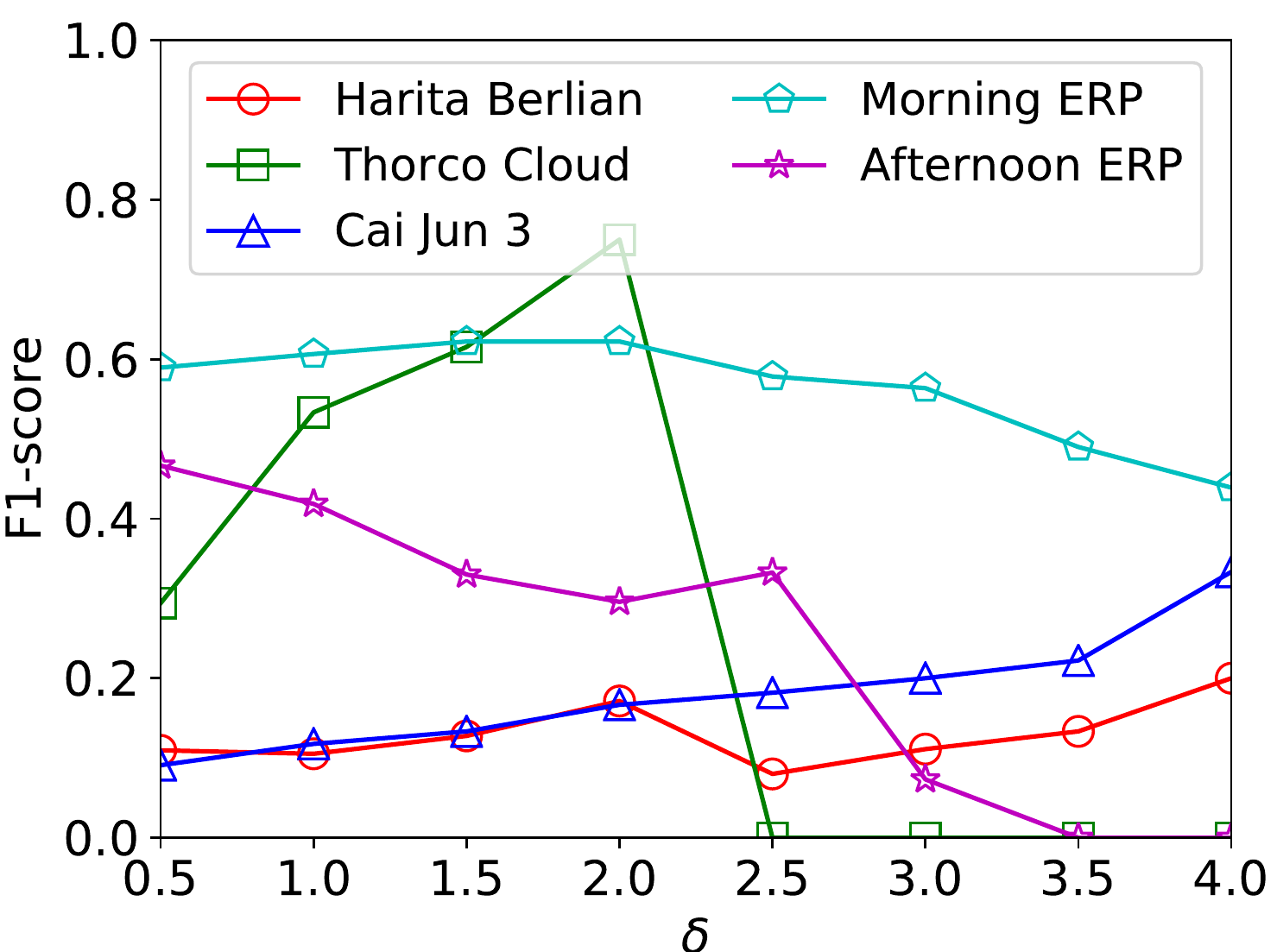}}
\subfigure[Query time vs.\ $\delta$]{
	\label{fig:param:querytime_delta}
	\includegraphics[width=0.235\textwidth]{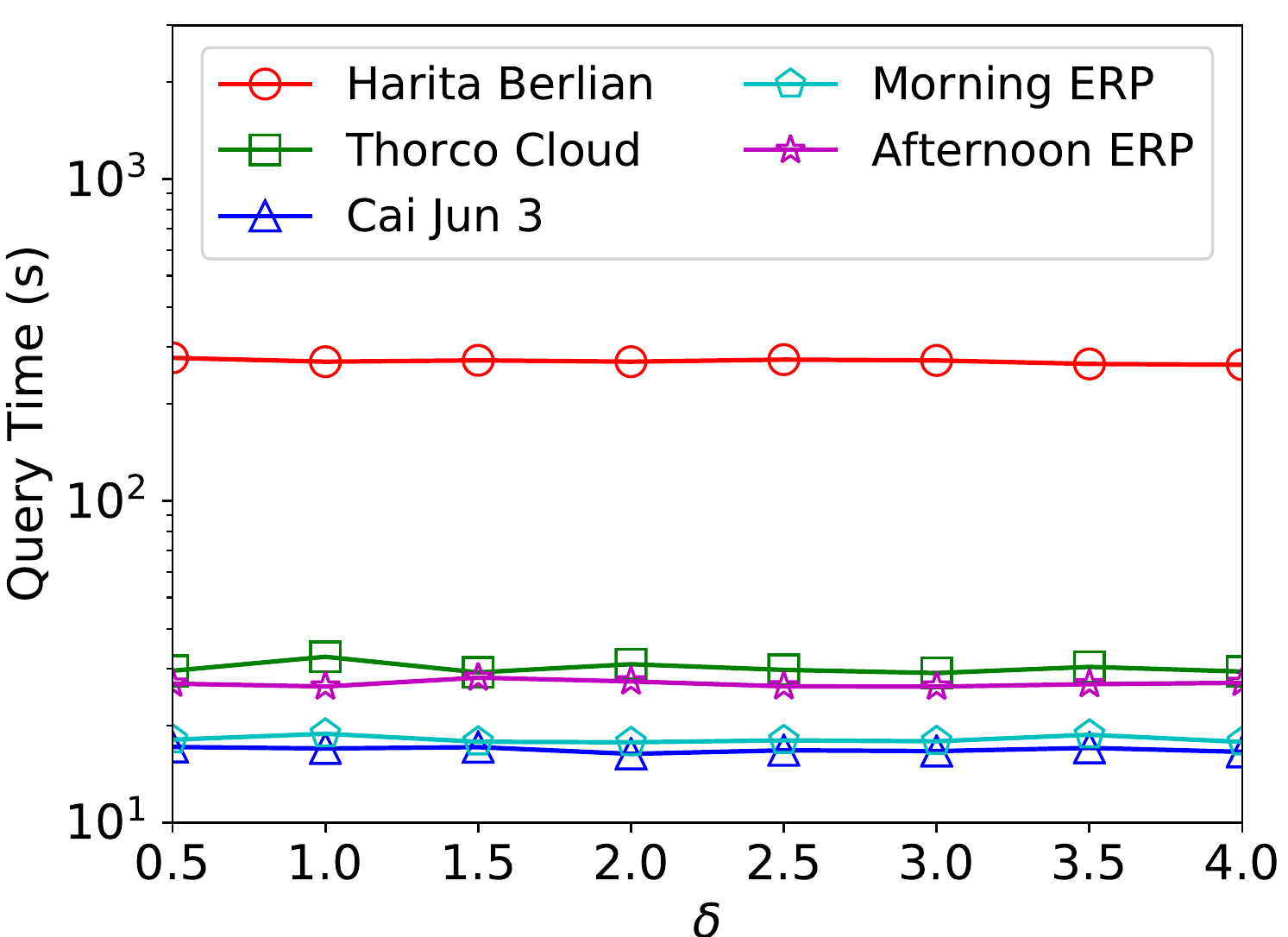}}
\subfigure[F1-score vs.\ $\tau$]{
	\label{fig:param:f1score_tau}
	\includegraphics[width=0.235\textwidth]{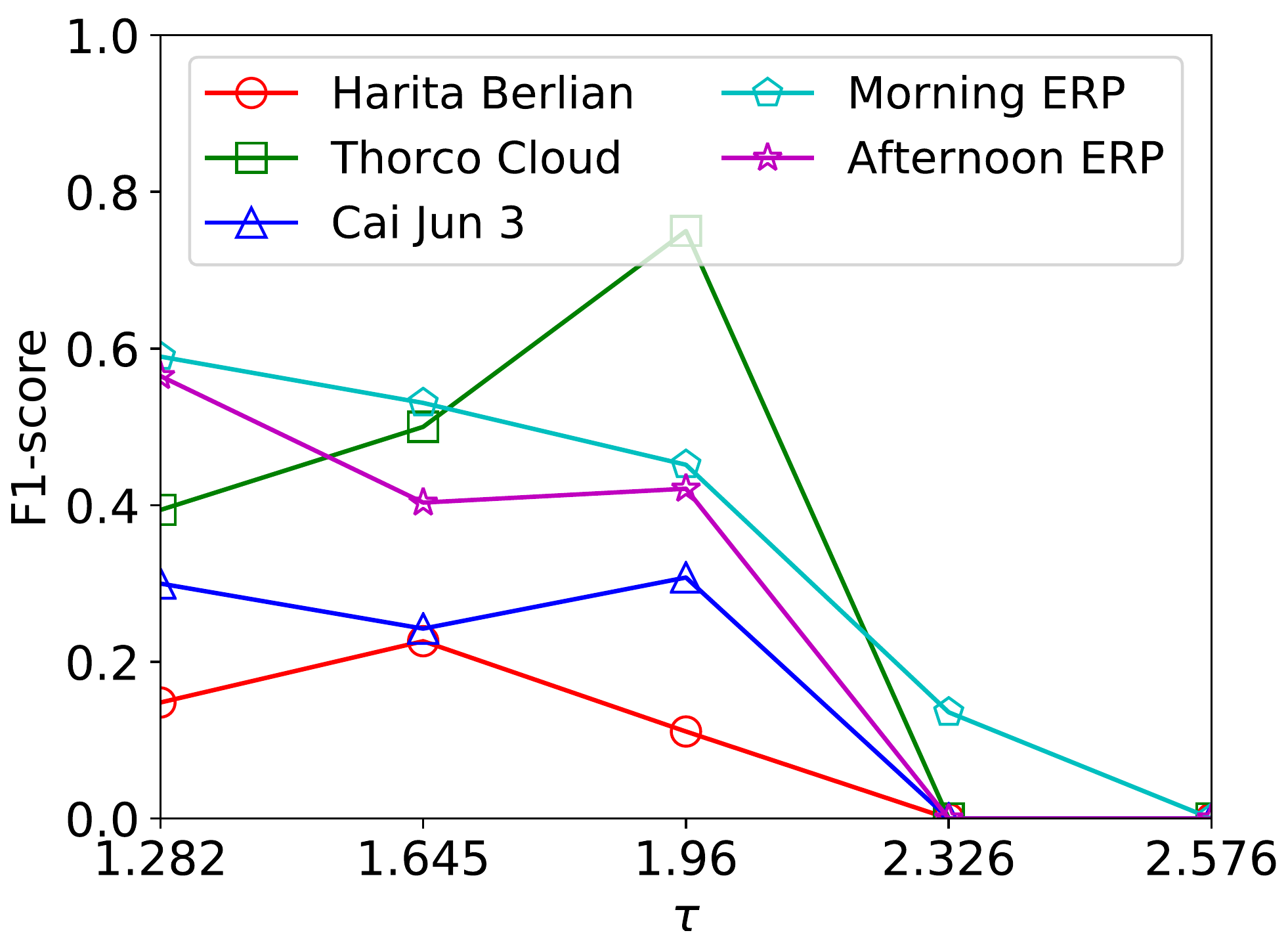}}
\subfigure[Query time vs.\ $\tau$]{
	\label{fig:param:querytime_tau}
	\includegraphics[width=0.235\textwidth]{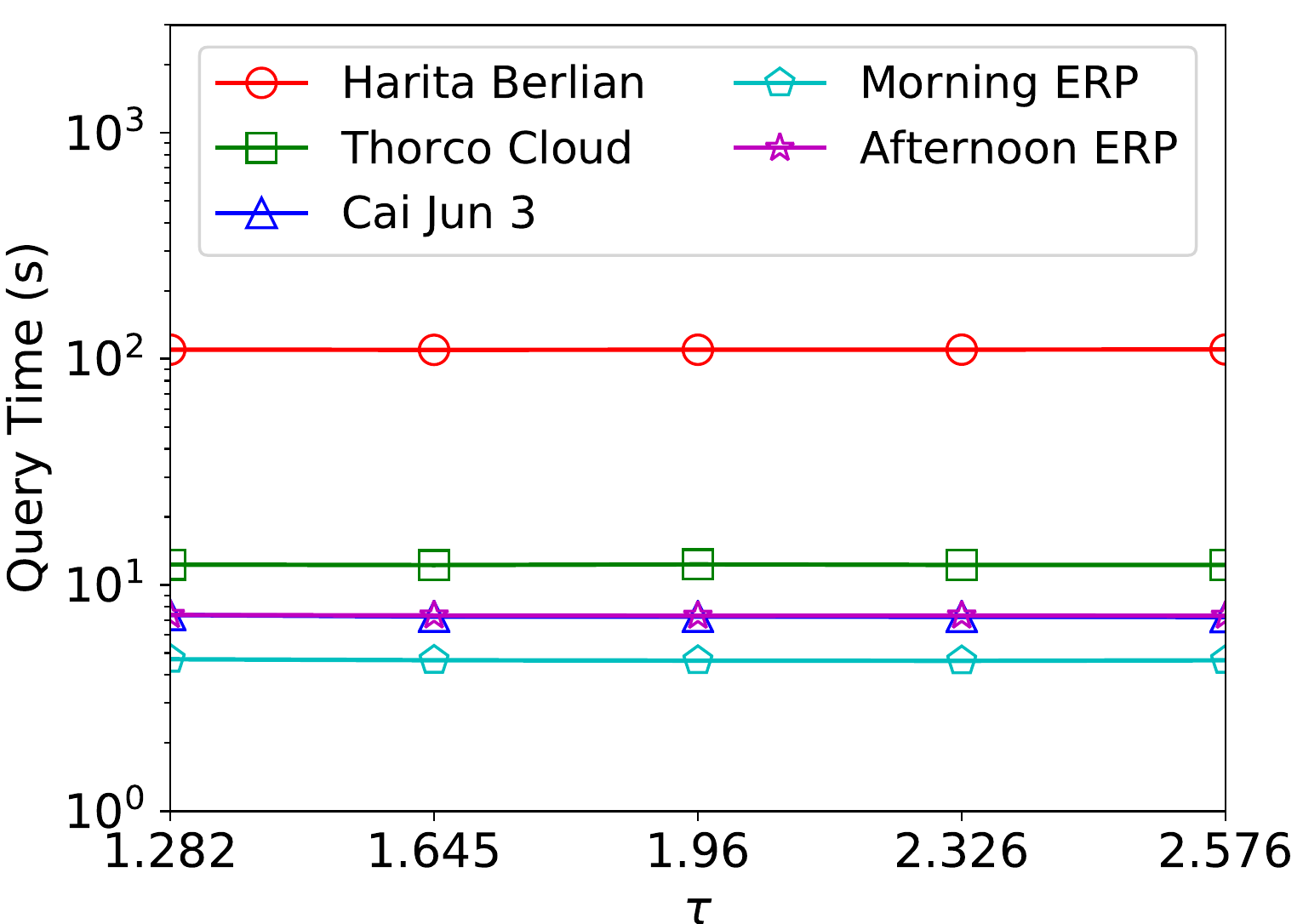}}
\caption{Effects of $\delta$ and $\tau$}
\label{fig:param}
\end{figure}

From Figure \ref{fig:param:f1score_delta}, the F1-score over $\delta$ generally has an \emph{inverse U} curve. This might be because when $\delta$ is small, increasing $\delta$ can increase the precision yet keep a high recall. However, when $\delta$ continues to increase, since some ground truths are missed, F1-score is decreased. 
From Figure \ref{fig:param:querytime_delta}, the query time is almost identical over different values of $\delta$ because the primary query time for obstacle detection is to retrieve the distinct $k$NNs of sub-trajectories. The impact of $\delta$ is not very sensitive to \textsf{DIOT}.

We then fix $\delta = 1.0$ and vary $\tau \in \{1.282,1.645,1.960,2.326,2.576\}$. The results are shown in Figures \ref{fig:param:f1score_tau} and \ref{fig:param:querytime_tau}. The pattern over various $\tau$ is similar to those of $\delta$, and the reason remains the same. 

%%%%%%%%%%%%%%%%%%%%%%%%%%%%%%%%%%%%%%%%%%%%%%%%%%%%%%%%%%%%%%%%%%%%%%%
%% Conclusions
%%%%%%%%%%%%%%%%%%%%%%%%%%%%%%%%%%%%%%%%%%%%%%%%%%%%%%%%%%%%%%%%%%%%%%%
\section{Conclusions}
\label{sect:conclusions}
In this paper, we study a new data mining problem of obstacle detection that has applications in many scenarios. 
We focus on the trajectory data and introduce a density-based definition for the obstacle. The proposed definition can approximately describe and quantize the relativity, significance, and support properties of obstacles. 
With this definition, we introduce a novel framework \textsf{DIOT} for obstacle detection and develop four insightful strategies for optimization. 
The experimental results on two real-life data sets demonstrate that \textsf{DIOT} enjoys superior performance yet captures the essence of obstacles.

%Our definition for the obstacle likeliness weight is a general form of convolution for the trajectory density function. We note that this type of definition can be generalized to other problems such as perturbing trajectories to preserve privacy, trajectory de-noising, etc. We are also looking forward to possible extensions on the augmented trajectory data such as the social network activities with temporal-spatial information. 

\subsubsection*{Acknowledgements.}
This research is supported by the National Research Foundation, Singapore under its Strategic Capability Research Centres Funding Initiative. Any opinions, findings and conclusions or recommendations expressed in this material are those of the author(s) and do not reflect the views of National Research Foundation, Singapore.

%*************************************************************************
% Bibliography
%*************************************************************************
\bibliographystyle{splncs04}
\bibliography{FullPaper}

\end{document}